# Fully Tunable On-Chip Meta-Generator for Multidimensional Poincaré Sphere mapping

Jing Luan[1,2,+], Tiange Wu[1,2,+], Shuang Zheng[1,2,*], Zhenyu Wan[1], Yuan Meng[4], Yijie Shen[3,5], Kaiyuan Wang[1,2], Deming Liu[1,2], Jian Wang[1,*] and Minming Zhang[1,2,*]

[1]School of Optical and Electronic Information and Wuhan National Laboratory for Optoelectronics, Huazhong University of Science and Technology, Wuhan 430074, Hubei, China

[2]National Engineering Research Center for Next Generation Internet Access System, Wuhan, Hubei 430074, China

[3]Centre for Disruptive Photonic Technologies, School of Physical and Mathematical Sciences & The Photonics Institute, Nanyang Technological University, Singapore 637371, Singapore

[4]Mechanical Engineering and Materials Science, Washington University in St Louis, St Louis, MO, 63130, USA

[5]School of Electrical and Electronic Engineering, Nanyang Technological University, Singapore 639798, Singapore

* zshust@hust.edu.cn, jwang@hust.edu.cn, and mmz@hust.edu.cn

**Abstract:**
The angular momentum of light can be elegantly mapped onto high-order Poincaré spheres, providing a powerful framework for describing structured light beams. Such beams have shown extraordinary potential across diverse applications, including high-capacity optical communications, precision metrology, and quantum information processing. While various methods exist for generating structured light beams, the dynamic synthesis and flexible control of arbitrary vectorial states on diverse, multidimensional Poincaré spheres still rely on bulky free-space optical components, posing significant challenges for scalability and integration. To date, a fully tunable solution implemented on a single photonic chip has yet to be realized. Here, we present the first fully tunable on-chip meta-generator capable of dynamically mapping arbitrary scalar, vectorial, and hybrid modes onto the full hierarchy of Poincaré spheres, and even extending to a high-dimensional Poincaré hypersphere within a four-dimensional Hilbert space. Our device is implemented on an eight-channel space-multiplexed multimode silicon photonic integrated circuit, where densely integrated mode multiplexers, amplitude–phase modulators, and an inverse-designed multimode meta-waveguide together enable compact, precise, and programmable control of structured light. The multimode meta-waveguide directly maps eight on-chip guided modes to orbital angular momentum (OAM), supporting broadband generation of high-purity OAM modes with diverse polarization states and topological charges. By simultaneously engineering amplitude, phase, polarization, and topological charge, we achieve full-field control over OAM mode bases, enabling fully tunable access to arbitrary scalar and vectorial states across more than eight distinct Poincaré spheres. This work represents a significant step toward reconfigurable on-chip manipulation of multidimensional Poincaré spheres, paving the way for advanced applications in optical communications, quantum photonics, and beyond.

## Introduction

Light with circular polarizations and helical phase fronts, carries spin angular momentum (SAM) and orbital angular momentum (OAM) respectively, representing two fundamental properties of light waves and photons. These two degrees of freedom (DoFs) have recently been developed as an important toolbox for describing tailored structured light beams with unique transverse shapes [1, 2]. The modern exploration of optical angular momentum indicates that optical vortex beams with helical phase fronts also carry OAM [3]. Beyond the optical phase vortices, another intriguing class of structured beams is the vectorial vortices, which are characterized by spatially varying polarization distributions or polarization singularities [4-6]. Notably, the SAM and OAM DoFs of cylindrical vector beams exhibit a physically non-separable nature, further enhancing their versatility and applications. The fundamental Poincaré sphere (PS) is commonly used to describe the polarization states of scalar light fields. Building upon this geometric representation, recent advancements have extended its framework to encompass more complex light structures. Orbital angular momentum (OAM) modes, vector modes, and their intricate interconnections can now be mapped onto the surfaces of corresponding unit spheres [7-10], including the OAM Poincaré sphere (OPS), the higher-order Poincaré sphere (HOPS), and the hybrid-order Poincaré sphere (HyOPS), providing a more comprehensive depiction of structured light fields. For instance, by replacing the orthogonal bases on the poles of the fundamental Poincaré sphere with OAM modes, the evolution between Laguerre-Gaussian (LG) and Hermite-Gaussian (HG) modes can be visualized on an analogous OPS [7]. The fundamental PS has also been extended to describe the higher-order polarization states of generalized vector vortex beam. By selecting different combinations of OAM charges for left- and right-circular polarization states at the sphere poles, more complex vector modes can be represented on modified spheres, such as the HOPS [8] or the HyOPS [9]. On a HOPS, the two poles correspond to opposite OAM states with orthogonal circular polarizations, while the equator represents cylindrical vector (CV) beams, with special cases being the azimuthally and radially polarized light fields. Alternatively, when circularly polarized OAM states carry different topological charges, the resulting hybrid-order Poincaré sphere can describe full Poincaré beams or C-point singularities [10]. The unique phase and intensity distributions of these PS beams make them highly valuable for a broad range of applications, including mode-division multiplexing, optical micromanipulation, microscopy and optical metrology, to name just a few.

The dynamic and flexible generation of arbitrary structured light fields on diverse PSs is crucial for unlocking their full potential in various applications [11–19]. So far, the dynamic generation and manipulation of full PS beams has largely relied on free-space optical setups, such as spatial light modulators and lens arrays [20–23]. While many metasurface- and fiber-based schemes have been ingeniously designed and successfully demonstrated, they are still limited to generating a narrow range of structured beams and fall short in offering dynamic tunability and full reconfigurability [24-31]. Thanks to their advantages in miniaturization, versatility, and compatibility with complementary metal-oxide semiconductor (CMOS) technology, photonic integrated circuits (PICs) have emerged as a highly attractive platform for generating structured light beams. Promising approaches utilizing subwavelength structures on photonic integrated waveguides have been demonstrated to generate specific types of structured beams [32–43]. For instance, compact optical vortex emitters based on microring gratings have been proposed to extract light confined in whispering gallery modes, producing free-space beams with well-controlled OAM [33–36]. However, these microring resonant structures are limited to emitting fixed polarization state distributions with narrow optical bandwidth and cannot be tuned. More recently, a series of meta-waveguides combining various functional subwavelength photonic architectures with diverse waveguide platforms have demonstrated exceptional capabilities in controlling guided electromagnetic waves [37–40]. Driven by advanced inverse-design algorithms, these meta-waveguides can exhibit impressive performance with enhanced multifunctionality and ultracompact device footprint [41, 42]. An alternative approach utilizing a programmable photonic mesh has also been demonstrated for the flexible and dynamic generation of structured light beams [43, 44]. Nevertheless, the ability to dynamically generate arbitrary structured light beams across various PSs, including PS, OPS, HOPS, and HyOPS, remains an unresolved challenge. In this scenario, a laudable goal would be to develop an ultracompact and fully tunable PS beam generator on photonic integrated platform.

In this work, we propose and experimentally demonstrate a fully tunable Poincaré sphere beam generator implemented on multimode silicon photonic integrated circuits. This device is composed of four two-mode multiplexers, sixteen amplitude-phase modulators, and an inversely-designed multimode meta-waveguide, enabling the emission of arbitrary structured light beams on diverse PSs, including fundamental PSs, OPSs, HOPSs, and HyOPSs. Inspired by the mapping between OAM and other physical dimensions, such as spin, amplitude, frequency, and time [45–47], we employ an inverse-designed photonic-crystal-like (PhC-like) meta-waveguide to achieve efficient multimode-to-OAM conversion within a waveguide platform. This meta-waveguide enables one-to-one conversion between eight space-multiplexed waveguide modes ($TE_0$, $TE_1$) and eight

polarization- and charge-diverse optical OAM modes (*x*/*y*-polarized $OAM_{\pm 1,\pm 2}$). Leveraging the orthogonal OAM mode basis provided by the meta-waveguide, arbitrary structured light fields on different PSs are flexibly synthesized and dynamically controlled by modulating the amplitude and phase profiles of the on-chip guided modes. This first single chip for dynamically generating all structured light states across more than eight distinct PSs will permit new advances on OAM generation and applications in integrated photonics realm.

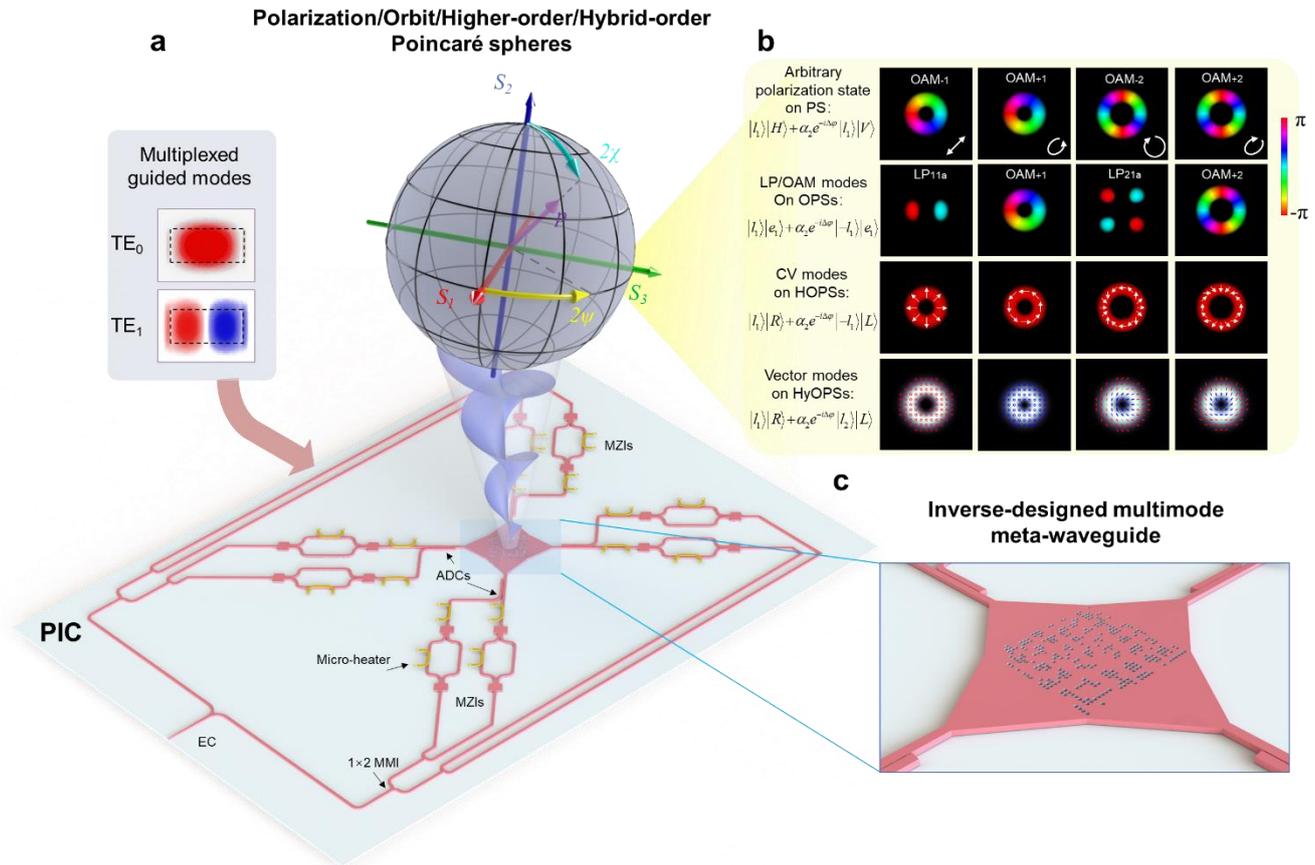

**Figure. 1 Device schematic diagram**. (**a**) Schematic of fully tunable Poincaré sphere beam generator based on silicon photonic integrated circuits (not to scale). PIC: photonic integrated circuits; EC: edge coupler; ADC: asymmetrical directional coupler for mode conversion and multiplexing; MZI: thermo-optic Mach-Zehnder interferometer for amplitude modulation; MMI: multimode interference coupler for optical power splitting. (**b**) Mode field profiles of typical modes on PSs, OPSs, HOPSs, and HyOPSs. (**c**) Inversely-designed multimode meta-waveguide, featuring a subwavelength surface structure etched onto a 2D multimode silicon waveguide, facilitates the conversion of in-plane eight waveguide modes into off-chip OAM modes.

## Results

### Multimode Integrated photonic circuit

Figure 1a illustrates the fully tunable Poincaré sphere beam generator based on an 8-channel multimode silicon photonic integrated circuit (PIC). The chip consists of four asymmetrical directional couplers (ADCs), eight tunable Mach-Zehnder interferometers (MZIs), eight phase modulators with thermo-optic microheaters, and an inversely-designed multimode meta-waveguide. This configuration enables the generation of all states across various PSs, as shown in Fig. 1b, including polarization PSs, OPSs, HOPSs, and HyOPSs. The light beam is first coupled into the chip through a high-efficiency edge coupler and is subsequently divided evenly into eight channels via a three-stage cascaded 1×2 splitter network. The amplitude and phase profiles of the input modes in each channel are precisely controlled using the amplitude and phase modulators. Following this, each ADC is used to multiplex fundamental transverse electric ($TE_0$) mode and higher-order transverse electric

($TE_1$) mode, allowing space multiplexing of four $TE_0$ modes and four $TE_1$ modes from four directions. The simulated eigenmode profiles, shown in Fig. 1a, reveal distinct electric field distributions for $TE_0$ and $TE_1$ modes. The $TE_0$ mode exhibits a simple optical field distribution, while the $TE_1$ mode features a more complex pattern, with its electric field divided into two lobes of equal amplitude and opposite phase. The inversely-designed multimode subwavelength meta-structure embedded on the surface of silicon waveguide (Fig. 1c) converts eight in-plane space-multiplexed guided modes ($TE_0$, $TE_1$) from four directions into eight polarization-/charge-diverse optical vortex modes ($x$/$y$-polarized $OAM_{\pm1, \pm2}$).

Notably, unlike our previous single-mode scheme, which is limited to the $TE_0$ mode, the proposed multimode meta-waveguide enables the conversion of higher-order waveguide modes, accommodating both $TE_0$ and $TE_1$ modes for enhanced versatility and functionality. This advancement doubles the number of polarization- and charge-diverse OAM modes that can be generated compared to the single-mode approach [39]. By leveraging the rich and orthogonal OAM mode basis provided by the meta-waveguide, the space-multiplexed waveguide modes can be further transformed into a broad range of complex structured light beams through precise control of the mode profiles across the eight input channels—a level of flexibility and diversity unattainable with previous design. As illustrated in Fig. 1b, the proposed scheme enables the flexible generation of arbitrary polarization states on the fundamental PSs, arbitrary linearly polarized (LP) modes on OPSs, and arbitrary vector modes on HOPSs or HyOPSs. For example, as shown in the second row of Fig. 1b, LP modes exhibit uniform polarization distributions while possessing spatially varying amplitude and phase profiles, which can be synthesized by superimposing two OAM modes with the same polarization state but opposite topological charges. CV modes, such as radially polarized (RP) and azimuthally polarized (AP) modes, exhibit spatially varying polarization states and can be generated by coherently combining two OAM modes with opposite topological charges and circular polarizations. Detailed mapping relationships between waveguide modes and structured light beams (OAM, LP, and vector modes) are given in Supplementary Section S4 (see Fig. S5 and Table S1).

**Inverse-designed multimode meta-waveguide**

Figure 2(a) shows the schematic of the proposed inversely-designed multimode meta-waveguide, enabling the conversion between eight space-multiplexed guided modes ($TE_0$, $TE_1$) and eight polarization-/charge-diverse OAM modes ($x$/$y$-polarized $OAM_{\pm1, \pm2}$). The design employs eight waveguide modes from four input ports. Due to the orthogonal polarization directions of the dominant electric fields in the $TE_0$ and $TE_1$ modes relative to the light propagation direction, $x$- or $y$-polarized OAM modes are generated when light propagates along the $y$- or $x$-direction, respectively. The sign and magnitude of the topological charge are determined by the propagation direction and the specific guided mode being excited. Figure 2(b) outlines the design process of the multimode meta-waveguide, which involves the formation and initialization of the holographic grating based on the principle of holographic superposition. This is followed by the transformation of the initial PhC-like structure into the optimized final design through inversely-design techniques.

The formation, principle, and theory foundation of the initial surface grating on the silicon waveguide are based on the holographic method, which can be described as follows. The coupled interference between the emitted OAM mode (propagating vertically) and the in-plane guided $TE_0$ mode (with the symmetric mode profile) gives rise to Grating 0 on the the waveguide surface. The electric fields of the target OAM mode and the guided $TE_0$ mode are expressed as:

$$E_{OAM} = O \exp(il\theta) \tag{1}$$

$$E_{waveguide\_TE_0} = W_{TE_0} \exp(i\beta_{TE_0} x) \tag{2}$$

where $O$ and $W$ are the amplitudes of the OAM mode and waveguide mode. $l$ and $\theta = \arctan^{-1}(y/x)$ represent the topological charge and azimuthal angle, respectively. $\beta_{TE_0}$ is the propagation constant of the waveguide mode. The interference of the guided mode and the OAM mode form the holographic grating on top of the waveguide, which can be given as [39]:

$$H_{\text{grating\_TE}_0} = \left| E_{\text{OAM}} + E_{\text{waveguide\_TE}_0} \right|^2 = O_{\text{TE}_0}^2 + W_{\text{TE}_0}^2 + 2O_{\text{TE}_0} W_{\text{TE}_0} \cos(\beta_{\text{TE}_0} x - l_{\text{TE}_0} \theta) \tag{3}$$

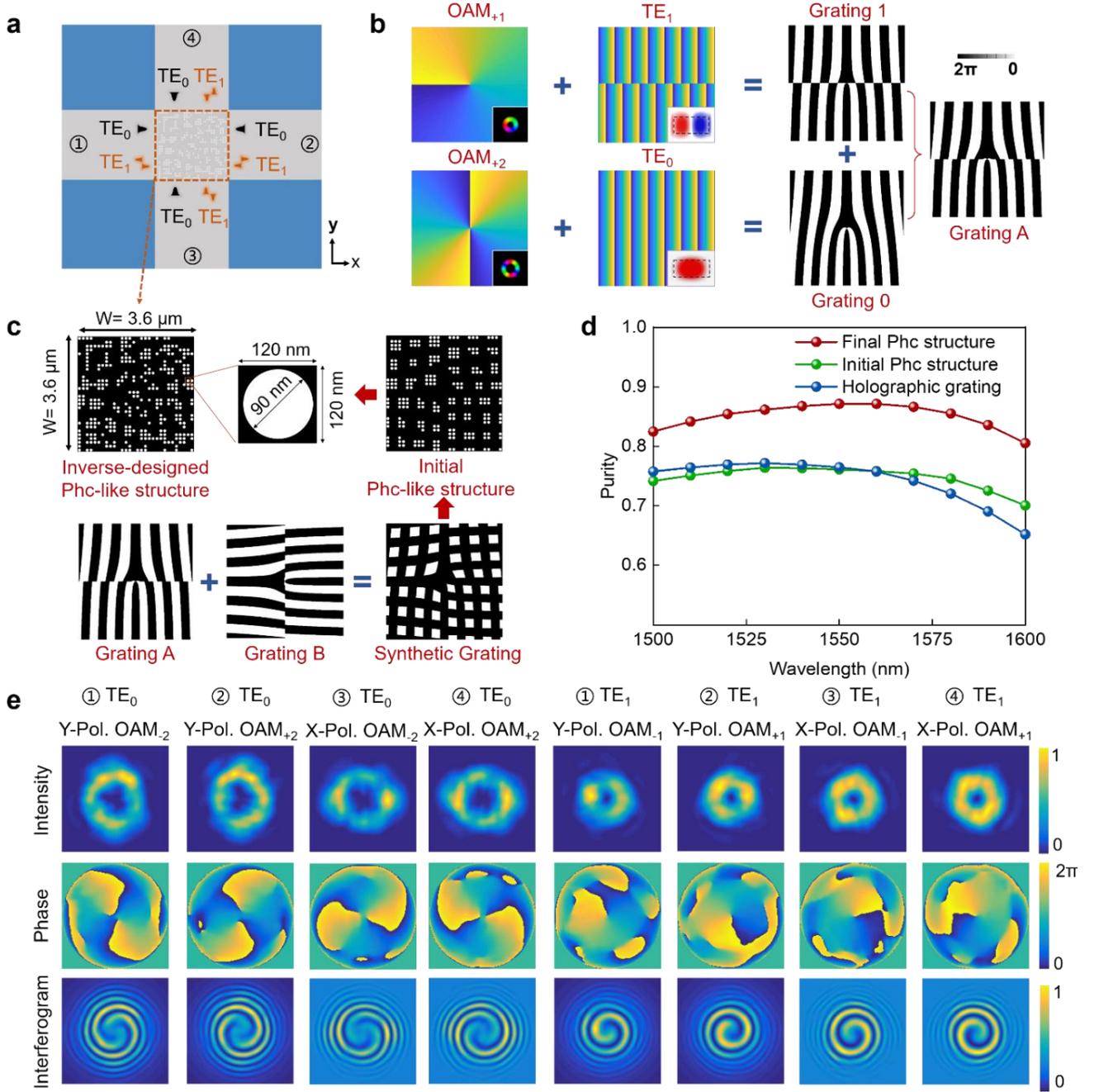

Figure. 2 **OAM mapping based on inversely-designed meta-waveguide**. (**a**) 2D subwavelength surface structure region with four input ports and two space-multiplexed modes (TE$_0$, TE$_1$). (**b**) Illustration of holographic method producing superposed fork grating. The superposed Grating A is synthesized by Grating 1 and 0. Fork grating 1 and 0 is superimposed by the phase profiles of OAM$_{+1}$ and TE$_1$, and the phase profiles of OAM$_{+2}$ and TE$_0$, respectively. (**c**) Zoom-in view of the optimized photonic-crystal-like (PhC-like) structure with inverse design method. The final holographic grating is superimposed by Grating A and its 90-degree rotated pattern B. The initial PhC-like structure is then obtained by digitized binary of the final holographic grating. (**d**) Calculated average purities of eight OAM modes (x/y-pol. OAM$_{\pm 1}$, x/y-pol. OAM$_{\pm 2}$) before and after optimization over wide wavelength range (length w: 3.6 µm; depth h: 130 nm). (**e**) Simulated intensity profiles, phase distributions, and interferograms of the generated polarization- and charge-diversity OAM modes at 1550 nm.

To accommodate practical fabrication techniques, a simple binary function is employed to define the grating's gray-scale distribution H (x, y). Specifically, H (x, y) = 0 when cos($\beta_0 x - l_0 \theta$) > 0, and H (x, y) = 1 otherwise. Based on the principle of holography, the guided wave propagating in the +x direction (from left to right) interacts with Grating 0 to generate a vortex beam with a topological charge of -2. Conversely, the guided wave propagating in the -x direction (from right to left) produces a vortex beam with a topological charge of +2.

For $TE_1$ mode with the asymmetric mode profile, the electric field can be written as

$$E_{waveguide\_TE_1} = \begin{cases} W_{TE_1} \exp(i\beta_{TE_1} x) & (y<0) \\ W_{TE_1} \exp(i(\beta_{TE_1} x + \pi)) & (y>0) \end{cases} \quad (4)$$

The distribution of the holographic grating can be given as:

$$H_{grating\_TE_1} = |E_{OAM} + E_{waveguide\_TE_1}|^2 = \begin{cases} O_{TE_1}^2 + W_{TE_1}^2 + 2OW_{TE_1} \cos(\beta_{TE_1} x - l_{TE_1}\theta) \\ O_{TE_1}^2 + W_{TE_1}^2 + 2OW_{TE_1} \cos(\beta_{TE_1} x - l_{TE_1}\theta + \pi) \end{cases} \quad (5)$$

Using the design principles of Grating 0, Grating 1 is developed to generate $OAM_{\pm 1}$ modes when excited by $TE_1$ modes propagating in the +x and -x direction, respectively.

It is important to note that for the strip waveguide with width of W = 3.6 μm, $TE_0$ and $TE_1$ modes have almost the same propagation constant $\beta_{eff} \approx 11.4$ μm$^{-1}$. To take advantage of mode multiplexing, Grating 0 and Grating 1 could be combined to form a new pattern Grating A, supporting multimode-to-OAM mapping. Thus, y-pol. $OAM_{-1}$, y-pol. $OAM_{+1}$, y-pol. $OAM_{-2}$ and y-pol. $OAM_{+2}$ modes could be generated when $TE_1$ mode (along +x direction), $TE_1$ mode (along -x direction), $TE_0$ mode (along +x direction) and $TE_0$ mode (along -x direction) are input. Similarly, Grating B can be generated by coupled interference between the vertically incident OAM modes (x-pol. OAM) and the in-plane guided modes (TE modes propagating along y direction). x-pol. $OAM_{-1}$, x-pol. $OAM_{+1}$, x-pol. $OAM_{-2}$ and x-pol. $OAM_{+2}$ modes can be converted by $TE_1$ mode (along -y direction), $TE_1$ mode (along +y direction), $TE_0$ mode (along -y direction) and $TE_0$ mode (along +y direction), respectively.

2D holographic grating can be synthesized by the superposition of grating A and B shown in Fig. 2(c). Then, the 2D holographic grating is discretized as a PhC-like array composed of 30×30 pixels with a footprint of 3.6×3.6 μm$^2$. Each pixel has a size of 120 nm ×120 nm and the shape of each pixel is a square with a circular silicon/silica hole at its center. The hole has a radius of 45 nm. The etching depth (h) of the circular could affect the OAM scattering efficiency and mode purity (Fig. S2). The initial PhC-like pixel array structure (at fixed h = 130 nm) can achieve average purity of more than 0.7 over wide wavelength range (1500-1600 nm, see Fig. S2b). The simulated far-field distributions for both 2D holographic grating and the initial PhC-like structure are shown in Fig. S3 (see Supplementary Section 2). The electric field distributions for the generated OAM modes are not complete due to imperfect mode conversion. The initial PhC-like structure is discretized by digitized binary of the final holographic grating and then optimized by the DBS inverse design algorithm (Fig. S1). Each pixel is filled by silicon or silica, corresponding to the logical "1" or "0" state respectively. The PhC-like subwavelength pixels can eliminate the random change of patterns of air holes and offer an outstanding RIE-lag-insensitive feature [48, 49]. The goal of our design is to obtain the optimized circular hole combinations (logical state arrangement) in the inverse design region to realize the maximum average phase purity of 8 emitted OAM modes. The FOM for the inverse design is defined as:

$$FOM = \frac{1}{8} \sum (P_{y,+1} + P_{y,+2} + P_{y,-1} + P_{y,-2} + P_{x,+1} + P_{x,+2} + P_{x,-1} + P_{x,-2}) \quad (6)$$

where $P_{y,+1}$, $P_{y,+2}$, $P_{y,-1}$, $P_{y,-2}$, $P_{x,+1}$, $P_{x,+2}$, $P_{x,-1}$ and $P_{x,-2}$ indicate the purity of vertically emitted y-Pol. $OAM_{+1}$, y-Pol. $OAM_{+2}$, y-Pol. $OAM_{-1}$, y-Pol. $OAM_{-2}$, x-Pol. $OAM_{+1}$, x-Pol. $OAM_{+2}$, x-Pol. $OAM_{-1}$ and x-Pol. $OAM_{-2}$ mode, respectively. For an ideal OAM emitter, the target FOM should be equal to 1. Figure S1(b) (in Supplementary Section S1) indicates the optimized 2D pixel array pattern and calculated FOM values of initial PhC-like structure and two iterations, respectively. As can be seen, the FOM increases from 0.75 to 0.84 after two iterations.

As depicted in Fig. 2d, eight high-purity OAM modes are successfully generated across a broad wavelength range spanning 1500–1600 nm. At the central wavelength of 1550 nm, the average mode purity reaches a peak value of 0.871, representing a 14.6% improvement over the holographic grating and the initial PhC-like structure. Detailed simulations of the mode purities are provided in Figs. S2(d–f) (see Supplementary Section S2).

The far-field intensity profiles, phase distributions, collinear interferograms of eight polarization- and charge-diversity OAM modes at 1550 nm are numerically simulated. To provide a clear visualization of the mapping relationship, the evolution from in-plane guided modes to out-of-plane OAM modes is illustrated in Fig. S4 (see Supplementary Section S3). As shown in Fig. 2(e), eight distinct polarization-/charge-diverse OAM modes are generated from on-chip guided modes, including $x$-Pol. $OAM_{-1}$, $x$-Pol. $OAM_{+1}$, $x$-Pol. $OAM_{-2}$, $x$-Pol. $OAM_{+2}$, $y$-Pol. $OAM_{-1}$, $y$-Pol. $OAM_{+1}$, $y$-Pol. $OAM_{-2}$, and $y$-Pol. $OAM_{+2}$. These modes are excited through specific guided mode-port configurations: $TE_1$ mode (Port 3), $TE_1$ mode (Port 4), $TE_0$ mode (Port 3), $TE_0$ mode (Port 4), $TE_1$ mode (Port 1), $TE_1$ mode (Port 2), $TE_0$ mode (Port 1), $TE_0$ mode (Port 2), respectively.

The intensity profiles exhibit characteristic annular structures, while the phase distributions reveal the expected spiral patterns. Specifically, the generated $OAM_{\pm 1}$ and $OAM_{\pm 2}$ modes demonstrate spiral phase distributions of $2\pi$ and $4\pi$, respectively. The collinear interferograms further confirm the correct topological charges, discernible through the counterclockwise or clockwise spiraling of the field intensity. Compared to the simulated results shown in Fig. S3 and Supplementary Section S2 for the holographic grating and the initial PhC-like structure, the far-field simulations in Fig. 2(e) exhibit more uniform electric field distributions. This improvement highlights an enhancement in mode purity achieved by the meta-waveguide design.

**Multidimensional Poincaré spheres Mapping**

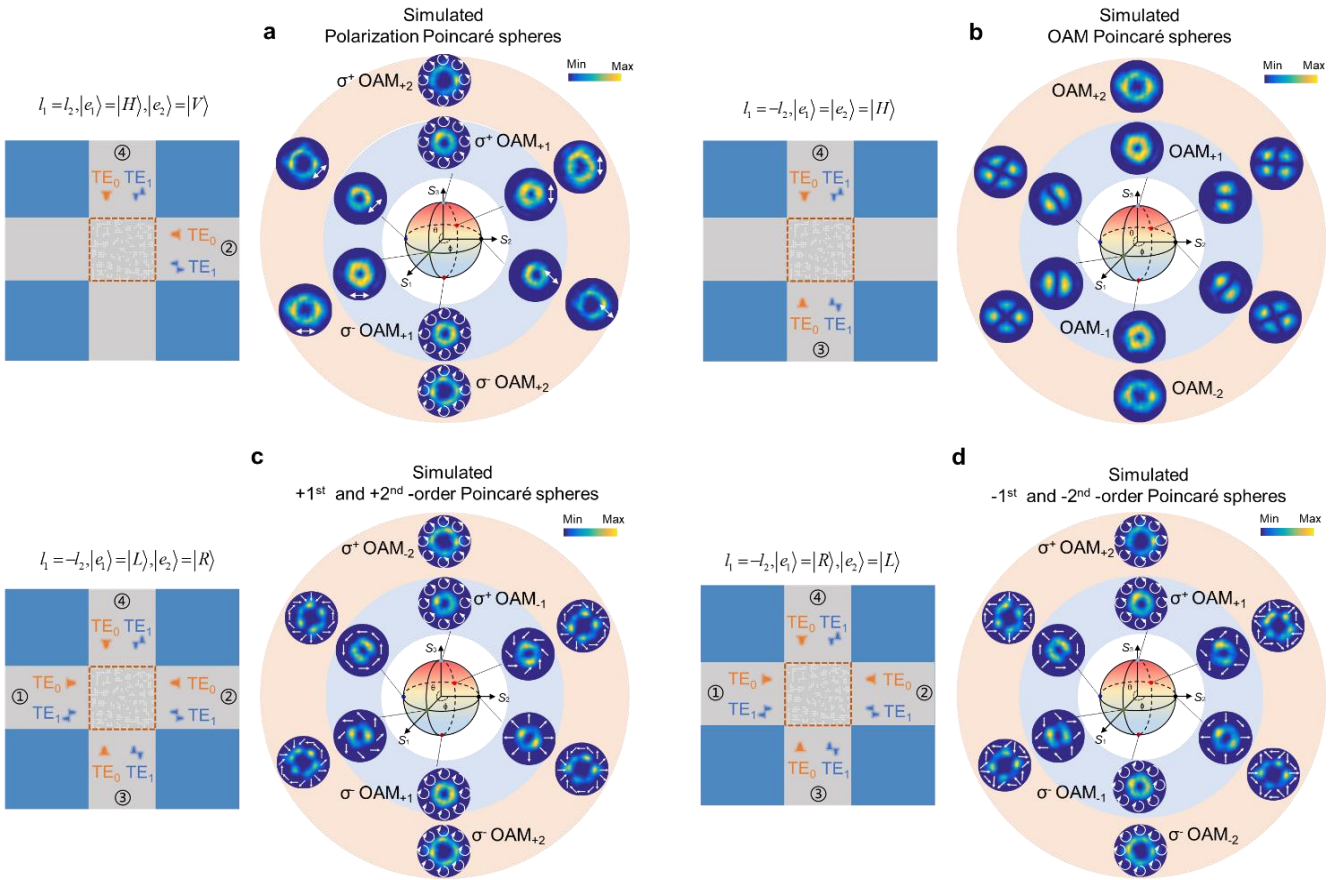

Figure. 3 **Controlled generation of high-dimensional vectorial states on different PSs.** (**a**) Generating vortex modes, with arbitrary uniform polarization state on the fundamental polarization PSs. (**b**) Typical OAM and LP modes on OAM PSs. (**c, d**) Typical CV modes on four HOPSs.

The proposed device offers the capability for dynamic mapping of guided modes to complex structured beams. Leveraging the OAM mode basis provided by the 2D-patterned meta-waveguide, on-chip guided modes can be flexibly mapped to various structured light beams, including arbitrarily polarized OAM modes on fundamental PSs, linearly polarized (LP) modes on OPSs, cylindrical vector (CV) modes on HOPSs, and more complex vector modes on HyOPSs. These structured beams on different PSs can be defined as,

$$|\psi\rangle = \alpha_1 |l_1\rangle |e_1\rangle + \alpha_2 e^{-i\Delta\varphi} |l_2\rangle |e_2\rangle \tag{7}$$

where $l_1$ and $l_2$ denote the OAM charges of the vortex beams with two scalar polarization states $|e_1\rangle$ and $|e_2\rangle$ respectively; the OAM charges and polarization states can be either identical or opposite, depending on the characteristics of the generated structured beams. $\alpha_{1,2}$ and $\Delta\varphi$ correspond to the amplitudes of the two vortex beams and their relative phase, respectively.

The scalar vortex beams $OAM_{+1,+2}$ can be generated with arbitrary polarization state on the fundamental PS when $l_1 = l_2 = +1/+2$ (Figs. 1b and 3a). Any point on the PS can be expressed as a coherent superposition of two orthogonal linear polarization states $|e_1\rangle$ and $|e_2\rangle$. By controlling the amplitude and phase between two orthogonal linearly polarized OAM modes derived from guided modes, arbitrary polarized OAM modes on the PS can be synthesized. For instance, $x$-polarized $OAM_{+1}$ mode is generated by the $TE_1$ mode of Port 4, while $y$-polarized $OAM_{+1}$ mode is derived from the $TE_1$ mode of Port 2. The coherent superposition of these two modes can form a new $OAM_{+1}$ mode with arbitrary scalar polarization states. Figure 3(a) shows the simulated optical fields for $OAM_{+1}$ and $OAM_{+2}$ modes, displaying six representative polarization states, including left- and right-circular polarization and four linear polarization states. These simulations confirm that the guided modes can be mapped to arbitrarily polarized OAM modes on the fundamental PS.

Similarly, an analogous OAM sphere is constructed to represent superpositions of left- and right-handed OAM modes, where the OAM charges satisfy $l_1 = -l_2 = +1/+2$, while maintaining an identical scalar polarization state. Any LP modes on OPSs can be generated by controlling the amplitude and phase profiles of the oppositely propagating guided modes. For example, $y$-polarized $OAM_{-1}$ is generated by the $TE_1$ mode of Port 1, while $y$-polarized $OAM_{+1}$ originates from the $TE_1$ mode of Port 3. By tuning the relative amplitude and phase values, the guided modes can be converted to arbitrary modes on the OPSs. As shown in Fig. 3(b), the relative phase of the superposition dictates the orientation of LP modes on the equator. Similar results for $OAM_{\pm 2}$ superpositions are observed in the brown region of Fig. 3(b).

Furthermore, the generation of first- and second-order CV beams can be achieved by controlling multiple guided modes. Figure 3(c) depicts two typical HOPSs, where the north and south poles represent left circularly polarized $OAM_{-l}$ and right circularly polarized $OAM_{+l}$, respectively. To synthesize a first-order CV beam on the HOPS, the phase relationship among four orthogonal OAM modes ($x$-$OAM_{\pm 1}$, $y$-$OAM_{\pm 1}$) must be precisely adjusted (see Supplementary Section S4). Detailed mapping relationships for CV modes are provided in Supplementary Table S1. In the inner blue region of Fig. 3(c), first-order CV beams with cylindrical symmetry in both polarization and phase distributions are located on the equator, including radially and azimuthally polarized beams. These beams exhibit circular intensity profiles and spatially anisotropic polarization distributions (see Supplementary Fig. S6). The outer brown region represents second-order CV beams synthesized by four orthogonal OAM modes ($x$-$OAM_{\pm 2}$, $y$-$OAM_{\pm 2}$). Simulated electric field components and corresponding phase distributions (see Supplementary Fig. S7) confirm that the polarization distributions align with theoretical predictions. More complex vector beams on the HyOPS can be realized by superimposing two scalar OAM modes with $l_1 \neq -l_2$, while ensuring opposite circular polarization states.

**Chip fabrication and measurement**

The chip is fabricated using a standard 220 nm-thick silicon photonic platform. Detailed information on the fabrication process and characterization is provided in Supplementary Section 5 (see Fig. S8). The fully packaged chip, integrated with a thermos-electric cooler (TEC), is shown in Fig. 1(a). All tuning components are wire-bonded to a carrier printed circuit board (PCB), and optical input/output is achieved using optical fiber arrays. The eighteen electrical tuning elements are controlled by a programmable control circuit (PCC) to enable dynamic reconfiguration. Figure 4(b) presents microscope images of the fabricated chip, showcasing the integration of an edge coupler, three-stage cascaded 1×2 splitters, eight thermo-optic MZIs, sixteen microheaters, four ADCs, and an inverse-designed meta-waveguide. Partial close-up images of the chip are provided in Figs. 4(c, d). Figure 4(e) shows a scanning electron microscope (SEM) image of the inverse-designed meta-waveguide, revealing a PhC-like pixelated pattern that matches the optimized design shown in Fig. 2(c). Additional SEM images of the edge waveguide coupler and the ADC are shown in Figs. 4(f, g). The edge coupler facilitates efficient fiber-to-chip coupling, while the ADC is responsible for (de)multiplexing the $TE_0$ and $TE_1$ modes. The transmission spectra of the fabricated ADC, shown in Fig. 4(h), demonstrate low insertion loss and crosstalk (< -20 dB) for both $TE_0$ and $TE_1$ modes, verifying the high-performance operation of the device.

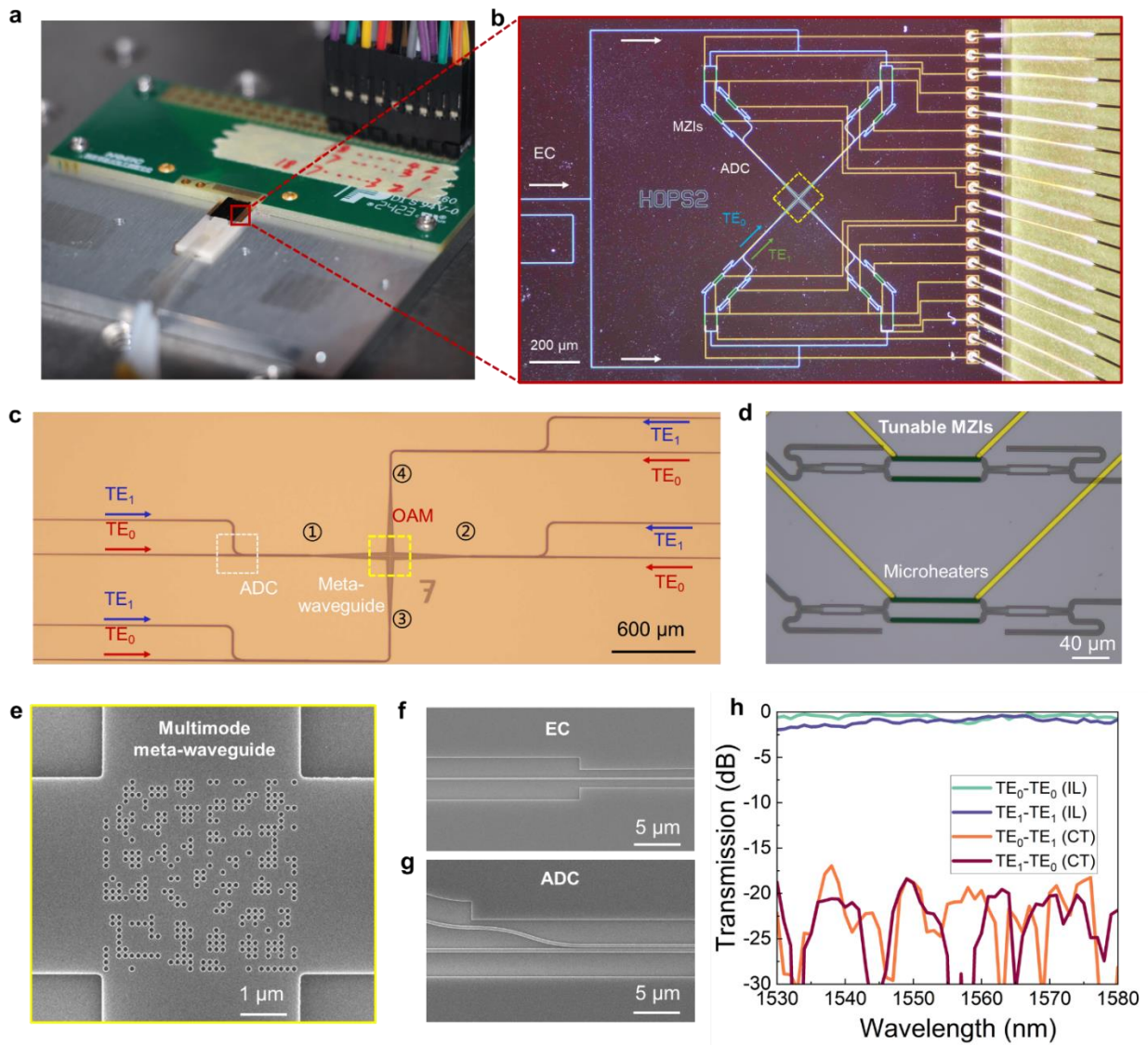

Figure 4. **Device fabrication and characterization.** (a) Microscopic image of the fully packaged chip, and (b) the fabricated silicon chip.

(c, d) Partial close-up images of the chip. (e) SEM image of the inverse-designed meta-waveguide with PhC-like pixelated pattern. (f, g) SEM image of the EC and ADC. (h) The measured transmission spectra of the ADC.

The experimental setup shown in Fig. S13 (see Supplementary Section S9) is used to characterize the performance of the fabricated chip. First, we measure the far-field intensity profiles and coaxial interferograms of eight polarization- and charge-diversity OAM modes (*x*-Pol. $OAM_{-1}$, *x*-Pol. $OAM_{+1}$, *x*-Pol. $OAM_{-2}$, *x*-Pol. $OAM_{+2}$, *y*-Pol. $OAM_{-1}$, *y*-Pol. $OAM_{+1}$, *y*-Pol. $OAM_{-2}$, and *y*-Pol. $OAM_{+2}$) at 1550 nm. As shown in Fig. 5(a), doughnut shaped intensity profiles and spiral interference patterns verify the helical phase fronts of the generated OAM modes. Topological charge orders can be recognized by the number of spiral arms and the spiral direction in the interferograms, which are consistent with the simulation results. To verify the polarization state of the generated OAM modes, we rotate the polarizer before the camera to record the power variation of *x*-Pol. $OAM_{-2}$ and y-Pol. $OAM_{+1}$ modes to analyze the characteristic of the polarization (see Fig. S10 in Supplementary Section S6).

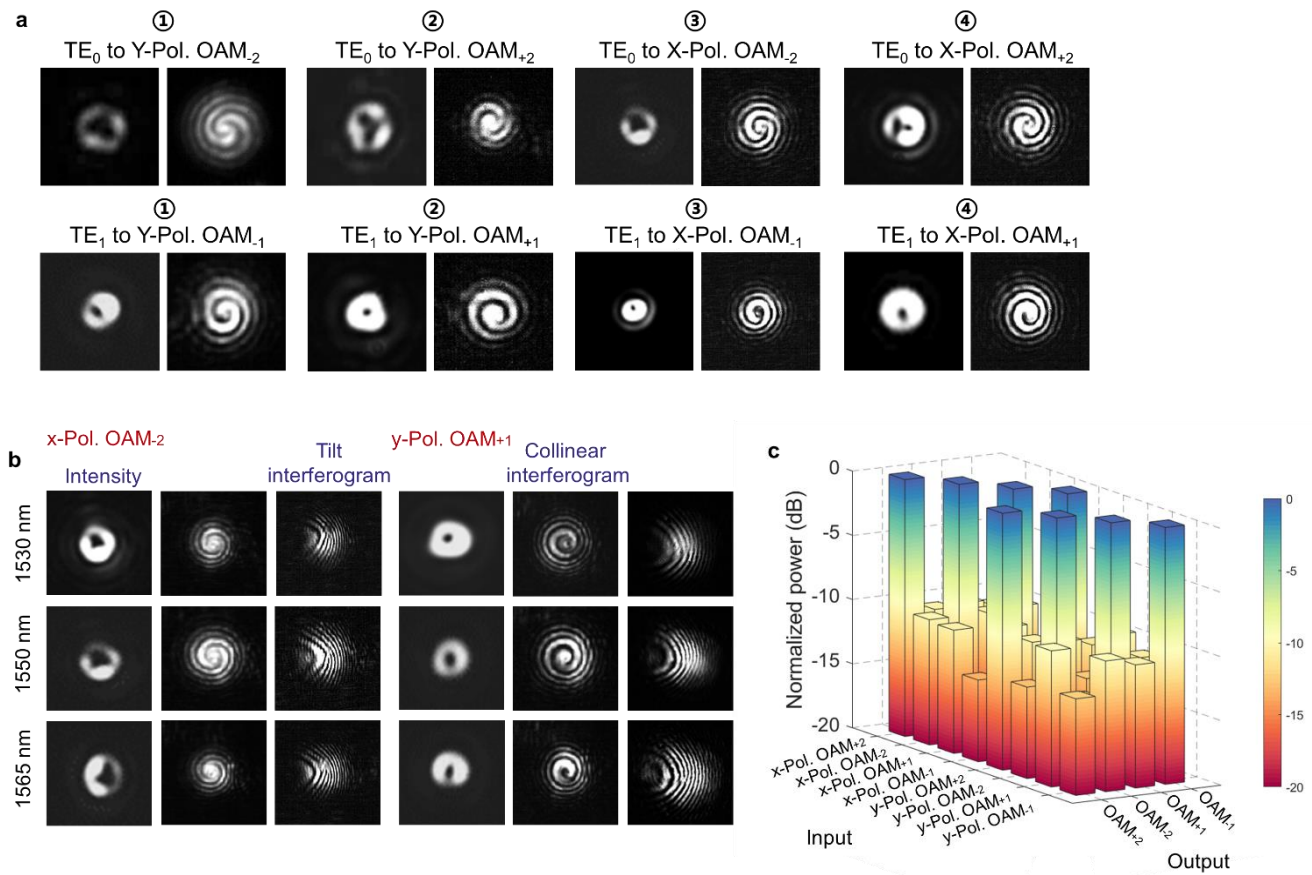

Figure 5. **Experimental measurements of eight polarization- and charge-diversity OAM modes**. (a) Measured far-field intensity profiles, collinear and tilt interferograms of the generated eight polarization- and charge-diversity OAM modes. (b) Measured far-field intensity profiles, collinear and tilt interferograms of the generated *x*-pol. $OAM_{-2}$ and *y*-pol. $OAM_{+1}$ modes at 1530, 1550, and 1565 nm for the meta-waveguide. (c) Measured histograms of the 8 × 4 crosstalk matrix for the meta-waveguide.

We also demonstrate the broadband generation of OAM modes from 1530 to 1560 nm. The measured far-field intensity profiles, collinear interferograms, tilt interferograms of *x*-Pol. $OAM_{-2}$ and *y*-Pol. $OAM_{+1}$ modes are shown in Fig. 5(b), respectively, at 1530, 1550 and 1565 nm. Hollow shaped intensity profiles, spiral collinear interference patterns, and fork-shaped tilt interferograms reveal good performance of the generated OAM modes. The measured additional results for the meta-waveguide are shown in Fig. S9. To measure the mode crosstalk of the generated OAM modes, four kinds of phase patterns are selectively loaded onto the spatial light modulator (SLM) to demodulate the $OAM_{+1}$, $OAM_{-1}$, $OAM_{+2}$, $OAM_{-2}$ modes into Gaussian-like beam, respectively. As shown in Fig. S11, the emitted OAM modes of the meta-waveguide can be demodulated to a series of bright spots at the beam center (Gaussian-like beam) when meeting an inverse spiral phase pattern, while remains the doughnut field profiles with other topological charges in the case of non-corresponding demodulation phase patterns. Figure 5(c) shows the recorded crosstalk matrices, which can evaluate the influence on each other for the generation of polarization- and charge-diversity OAM modes for the multimode meta-waveguide at 1550 nm. As can be seen from the measured histograms, the average crosstalk values are less than −10 dB.

**Tunable generation of Poincaré sphere beams**

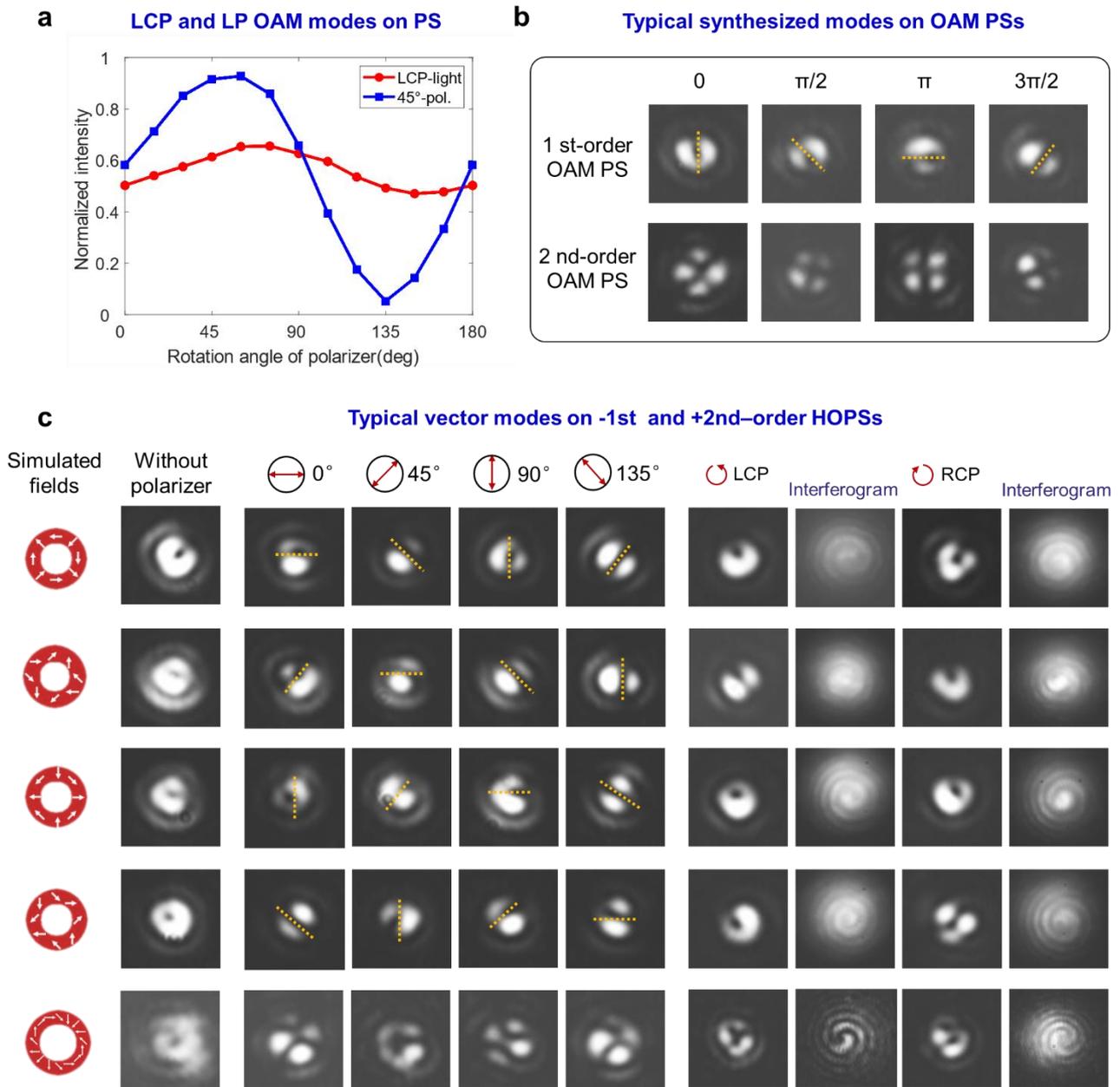

Figure 6. **Tunable generation of diverse PS beams**. (a) Typical OAM modes with LCP and LP states on polarization PSs. (b) Typical OAM and LP modes on two OAM PSs. (c) Typical vector modes on -1st and +2nd-order HOPSs.

The generated OAM mode basis can serve as a versatile platform for synthesizing all states across fundamental PSs, OPSs, and HOPSs and HyOPSs. For example, arbitrary polarization states on a fundamental Poincaré sphere can be generated by selecting two orthogonally polarized OAM modes ($x$-pol. OAM and $y$-pol. OAM), which can be derived from the guided modes in Ports 1 and 3 or Ports 2 and 4. To validate this capability, we experimentally demonstrate the generation of two representative polarized OAM beams. Figure 6(a) shows the measured normalized intensity versus the rotation angle of a polarizer. The sinusoidal blue curve corresponds to a 45°-polarized linearly polarized $OAM_{-1}$ beam, confirming its successful generation. Similarly, the red curve represents the normalized intensity as a function of the polarizer angle for a left-handed circularly polarized (LCP) $OAM_{-1}$ beam, illustrating precise control over polarization states. These results confirm that the fabricated device can generate $OAM_{\pm 1}$ and $OAM_{\pm 2}$ modes with arbitrary uniform polarization states across fundamental PSs.

The fabricated chip can also generate arbitrary states on analogous OAM spheres. Distinctive from fundamental PSs, the poles of the OAM spheres correspond to two OAM modes with identical polarization states but opposite topological charges, such as $OAM_{+1}$ and $OAM_{-1}$. In the experiment, $y$-polarized $OAM_{-1}$ is converted from the $TE_1$ mode of Port 1, and $y$-polarized $OAM_{+1}$ is derived from the $TE_1$ mode of Port 3, while the other six channels are deactivated by tuning the MZIs. By adjusting the thermo-optic phase shifters of these two channels, four typical LP modes on the first-order OAM Poincaré sphere are generated in Fig. 6(b). As shown in Fig. 3(b), the relative phase of the superimposed modes determines the orientation of the LP modes along the equator. Similar results for the superposition of $OAM_{\pm 2}$ modes are also observed, demonstrating the flexibility of the device in generating arbitrary states on OAM PSs.

The chip also demonstrates the capability to flexibly generate complex CV beams on different HOPSs. As shown in Fig. 6(c), the chip is employed to emit typical vector modes on the -1st and +2nd-order HOPSs. The first row illustrates the measured intensity profiles of a vector mode located on the equator of -1st-order HOPS, where a polarizer with varying orientation angle is used to verify the polarization state of the emitted light fields. When the light passes through the polarizer with orientation angle of 0° and 90°, two bright spots appear along the vertical and horizontal directions, respectively. A similar phenomenon is observed for polarizer orientation angles of 45° and 135°, confirming that the measured polarization distributions align well with the target polarized CV beam. To further validate the polarization state of generated CV beam, circularly polarized Gaussian beams are coaxially interfered with the light fields. A first-order CV beam can be described as a superposition of a vortex mode with RCP $OAM_{-1}$ and a vortex mode with LCP $OAM_{+1}$. When the generated light field is interfered with a reference LCP (RCP) Gaussian beam and analyzed through a circularly polarized filter composed of a quarter-wave plate (QWP) and a polarizer, the resulting interferograms exhibit helical interference fringes. These fringes verify the presence of LCP $OAM_{+1}$ and RCP $OAM_{-1}$ components, as shown in Fig. 6(c).

The experiment also investigates the generation of CV beams with other polarization states. As depicted in Fig. 6(c), CV beams with four distinct states located on the equator of the sphere are presented, corresponding to polarization orientation angles $\varphi_0 = 0°, 45°, 90°,$ and $135°$. After passing through the polarizer, the doughnut-shaped intensity profiles of the CV beams transform into two-lobed patterns. The orientation angles of the dark fringes in these patterns confirm the polarization states of the generated CV beams. Additionally, the chip achieves the generation of a typical vector mode on the +2nd-order HOPS, featuring a more intricate polarization vector distribution. These results further showcase the versatility of the chip in generating structured light fields with complex polarization characteristics. It is worth noting that more than the states on the equator and longitude, any point on the surface of PSs can be achieved by adjusting the relative amplitude and phase difference of four input guided waves, despite of these specific cases adopted in this work for demonstration.

## Discussion and Conclusion

In summary, we have successfully demonstrated a fully tunable Poincaré sphere beam generator on a silicon photonic chip. The device integrates an edge coupler, three-stage cascaded 1×2 splitters, eight tunable MZIs with thermo-optic phase shifters, sixteen microheaters, four ADCs, and an inversely-designed multimode meta-waveguide. Notably, leveraging the multimode meta-waveguide for eight-channel mode conversion and eight-channel amplitude-phase modulation, our platform enables dynamic generation of structured light beams spanning more than eight distinct PSs. Experimentally, we have demonstrated the versatility of the device by generating a range of structured beams, including arbitrary polarization states on the fundamental PS, four LP modes on the first- and second-order OAM PSs, and CV modes on the first- and second-order HOPSs.

The successful realization of these diverse structured light beams validates our chip design and implementation, paving the way for high-performance and reconfigurable photonic integrated circuits for practical applications.

A comparison of our work with state-of-the-art on-chip structured light beam generators is summarized in Table S2 (see Supplementary Section S10). By leveraging an inversely-designed multimode meta-waveguide that supports both fundamental and higher-order modes, our approach facilitates broadband and efficient conversion between eight guided modes and eight orthogonal OAM modes. Further empowered by a thermo-optically tunable silicon photonic platform, our device facilitates the generation of arbitrarily structured beams across more than eight distinct PSs, surpassing previously reported works in the number and variety of modes [28, 37, 39, 40, 42-44]. Moreover, the integration of amplitude and phase modulators significantly enhances dynamic reconfigurability.

Finally, several enhancements could further expand the impact of this work. The device footprint can be significantly reduced by leveraging inversely-designed mode-division multiplexed photonic circuits, enabling ultra-dense on-chip integrated PS beam generator [48–50]. The switching speed, currently limited by the thermo-optic phase shifters, could be dramatically improved to the nanosecond scale by incorporating faster modulators, such as PN junction phase shifters [51, 52]. The integration of on-chip laser sources would enhance both scalability and compactness, paving the way for fully integrated, high-performance structured light beam generation systems. Since our chip can simultaneously manipulate vector beams on two distinct PSs, it serves as a powerful tool for the reconfigurable and dynamic generation of optical skyrmions, providing a versatile platform for customizing topological light fields on demand [53-56]. In addition, the demonstrated generator can be further developed for dynamical generation and reconfiguration of high-dimensional superposition states in a four-dimensional Hilbert space [57, 58].

## Methods

**Chip fabrication and assembly process**

As indicated in Fig. S8, the photonic chip is fabricated and experimentally demonstrated on standard 220 nm silicon-on-insulator (SOI) wafer. More details can be found in Supporting information, Section S5. The chip is mounted onto a copper chuck using thermal epoxy for efficient heat dissipation. Wire bonding is employed to establish electrical connections between the chip and the printed circuit board (PCB). Low-loss, broadband fiber-to-chip coupling is achieved via inverse-taper edge couplers, facilitating efficient optical interfacing. A fiber array with a 127-μm pitch is precisely aligned and fixed to the on-chip edge coupler array. To ensure thermal stability, a thermo-electric cooler (TEC) is integrated, maintaining the chip at a controlled ambient temperature.

**Device simulation**

The finite-difference time-domain (FDTD) method is used to perform the physical simulations of the meta-waveguide, ADC, splitter, and optical MZI. The finite element method (FEM) method is utilized to calculate the waveguide mode profiles.

## Data availability

The data that support the findings of this study are included in the article and its supplementary information. Other data are available from the corresponding author upon request.

## References

1. Franke-Arnold, S., Allen, L. & Padgett, M. Advances in optical angular momentum. Laser Photonics Rev. **2**, 299–313 (2008).
2. Yao, A. M. & Padgett, M. J. Orbital angular momentum: origins, behavior and applications. Adv. Opt. Photon. **3**, 161–204 (2011).
3. L. Allen, M. W. Beijersbergen, R.J. C. Spreeuw, & J. P. Woerdman. Orbital angular momentum of light and the transformation of Laguerre-Gaussian laser modes. Phys. Rev. A **45**, 8185–8189 (1992).

4. Zhan, Q. W. Cylindrical vector beams: from mathematical concepts to applications. Adv. Opt. Photon. **1**, 1–57 (2009).
5. Shen, Y., et al. Optical vortices 30 years on: OAM manipulation from topological charge to multiple singularities. Light Sci. Appl. **8**, 90 (2019).
6. Forbes, A., De Oliveira, M. & Dennis, M. R. Structured light. Nat. Photonics **15**, 253–262 (2021).
7. Padgett, M. J. & Courtial, J. Poincare-sphere equivalent for light beams containing orbital angular momentum. Opt. Lett. **24**, 430–432 (1999).
8. Milione G., Sztul H. I., Nolan D. A. & Alfano R. R. Higher-order poincaré sphere, stokes parameters, and the angular momentum of light. Phys. Rev. Lett. **107**, 053601 (2011).
9. Yi, X. et al. Hybrid-order Poincaré sphere. Phys. Rev. A **91**, 023801 (2015).
10. Arora, G., Ruchi & Senthilkumaran, P. Full Poincare beam with all the Stokes vortices. Opt. Lett. **44**(22): 5638-5641 (2019).
11. Wang, J. et al. Terabit free-space data transmission employing orbital angular momentum multiplexing. Nat. Photonics **6**, 488–496 (2012).
12. Bozinovic, N. et al. Terabit-scale orbital angular momentum mode division multiplexing in fibers. Science **340**, 1545–1548 (2013).
13. Willner, A. E. et al. Optical communications using orbital angular momentum beams. Adv. Opt. Photon. **7**, 66–106 (2015).
14. Lavery, M.P., Speirits, F.C., Barnett, S.M. & Padgett, M.J. Detection of a spinning object using light's orbital angular momentum. Science **341**, 537–540 (2013).
15. Mair, A., Vaziri, A., Weihs, G. & Zeilinger, A. Entanglement of the orbital angular momentum states of photons. Nature **412**, 313–316 (2001).
16. Fang, X., Ren, H. & Gu, M. Orbital angular momentum holography for high-security encryption. Nat. Photonics **14**, 102–108 (2020).
17. Ren, H. R. et al. Complex-amplitude metasurface-based orbital angular momentum holography in momentum space. Nat. Nanotechnol. **15**, 948–955 (2020).
18. Fang, L., Wan, Z., Forbes, A. and Wang, J. Vectorial Doppler metrology. Nat. Commun. **12**, 4186 (2021).
19. Cheng, M., Jiang, W., Guo, L., Li, J. & Forbes, A. Metrology with a twist: probing and sensing with vortex light. Light Sci. Appl. **14**, 4 (2025).
20. Chen, S. et al. Generation of arbitrary cylindrical vector beams on the higher order Poincaré sphere. Opt. Lett. **39**, 5274-5276 (2014).
21. Naidoo, D. et al. Controlled generation of higher-order Poincaré sphere beams from a laser. Nat. Photonics **10**, 327–332 (2016).
22. Liu, Z. et al. Generation of arbitrary vector vortex beams on hybrid-order Poincaré sphere. Photon. Res. **5,** 15–21 (2016).
23. Sheng, L. et al. Highly efficient generation of arbitrary vector beams with tunable polarization, phase, and amplitude. Photon. Res. **6**, 228-233 (2018).
24. Yue, F. et al. Vector vortex beam generation with a single plasmonic metasurface. ACS Photon. **3**, 1558-1563 (2016).
25. Guo, X., Ding, Y., Chen, X., Duan, Y. & Ni, X. Molding free-space light with guided wave-driven metasurfaces. Sci. Adv. **6**, eabb4142 (2020).
26. Liu, M. et al. Broadband generation of perfect Poincaré beams via dielectric spin-multiplexed metasurface. Nat. Commun. **12**, 2230 (2021).
27. Dorrah, A. H., & Capasso, F. Tunable structured light with flat optics. Science, **376**, 1-11 (2022).
28. Ji, J. et al. Metasurface-enabled on-chip manipulation of higher-order Poincaré sphere beams. Nano Lett. **23** (7), 750-2757 (2023).
29. Lin, D. et al. Reconfigurable structured light generation in a multicore fibre amplifier. Nat. Commun. **11**, 3986 (2020).
30. Li, C. et al. Metafiber transforming arbitrarily structured light. Nat. Commun. **14**, 7222 (2023).
31. Feng, L. P. et al. All-fiber generation of arbitrary cylindrical vector beams on the first-order Poincaré sphere. Photon. Res. **8**, 1268-1277 (2020).
32. Su, T. H. et al. Demonstration of free space coherent optical communication using integrated silicon photonic orbital angular momentum devices. Opt. Express **20**, 9396-9402 (2012).
33. Cai, X. L. et al. Integrated compact optical vortex beam emitters. Science **338**, 363–366 (2012).
34. Miao, P. et al. Orbital angular momentum microlaser. Science **353**, 464–467 (2016).
35. Forbes, A., Mkhumbuza, L. & Feng, L. Orbital angular momentum lasers. Nat. Rev. Phys. **6**, 352–364 (2024).
36. Shao, Z. et al. Spin-orbit interaction of light induced by transverse spin angular momentum engineering. Nat. Commun. **9**, 926 (2018).
37. Xie, Z. W. et al. Ultra-broadband on-chip twisted light emitter for optical communications. Light.: Sci. Appl. **7**, 18001 (2018).
38. Zhou, N. et al. Generating and synthesizing ultrabroadband twisted light using a compact silicon chip. Opt. Lett. **43**, 3140-3143 (2018).
39. Zhou, N. et al. Ultra-compact broadband polarization diversity orbital angular momentum generator with $3.6 \times 3.6$ μm$^2$ footprint. Sci. Adv. **5**, eaau9593 (2019).

40. Song, H. et al. Demonstration of generating a 100 Gbit/s orbital-angular-momentum beam with a tunable mode order over a range of wavelengths using an integrated broadband pixel-array structure. Opt. Lett. **46**, 4765-4768 (2021).
41. Meng, Y. et al. Optical meta-waveguides for integrated photonics and beyond. Light Sci. Appl. **10**, 235 (2021).
42. Alexander, D. et al. Inverse Design of Optical Vortex Beam Emitters. ACS Photonics **10**, 803–807 (2023).
43. Bütow, J. et al. Generating free-space structured light with programmable integrated photonics. Nat. Photonics **18**, 243–249 (2024).
44. Lu, K., Chen, Z., Chen, H. et al. Empowering high-dimensional optical fiber communications with integrated photonic processors. Nat. Commun. **15**, 3515 (2024).
45. Paul, B., Minho, K., Connor, R. & Hui, D. High fidelity detection of the orbital angular momentum of light by time mapping. New J Phys **15**, 113062 (2013).
46. Fang, L., Wang, H., Liang, Y., Cao, H., & Wang, J. Spin-orbit mapping of light. Phys. Rev. Lett. **127**, 233901 (2021).
47. Khan, M. I. W. et al. A 0.31-THz orbital-angular-momentum (OAM) wave transceiver in CMOS with bits-to-OAM mode mapping. IEEE Journal of Solid-State Circuits, **57**, 1344-1357 (2022).
48. Lu, L. L. Z. et al. Inverse-designed single-step-etched colorless 3 dB couplers based on RIE-lag-insensitive PhC-like subwavelength structures. Opt. Lett. **41**, 5051–5054 (2016).
49. Chang, W. J. et al. Ultracompact dual-mode waveguide crossing based on subwavelength multimode-interference couplers. Photonics Res. **6**, 660–665 (2018).
50. Liu, Y. J. et al. Arbitrarily routed mode-division multiplexed photonic circuits for dense integration. Nat. Commun. **10**, 3263 (2019).
51. Li, M. F. et al. Silicon intensity Mach–Zehnder modulator for single lane 100 Gb/s applications. Photon. Res. **6**, 109-116 (2018).
52. He, M. et al. High-performance hybrid silicon and lithium niobate Mach–Zehnder modulators for 100 Gbit s$^{-1}$ and beyond. Nat. Photonics **13**, 359–364 (2019).
53. He, T. et al. Optical skyrmions from metafibers with subwavelength features. Nat. Commun. **15**, 10141 (2024).
54. Lin, W. B. et al. On-chip optical skyrmionic beam generators. Optica **11**, 1588-1594 (2024).
55. Shen, Y. J. et al. Topological bimeronic beams. Opt. Lett. **46**, 3737-3740 (2021).
56. Shen, Y. et al. Optical skyrmions and other topological quasiparticles of light. Nat. Photonics **18**, 15-25 (2024).
57. Zhang, Z. et al. Spin–orbit microlaser emitting in a four-dimensional Hilbert space. Nature **612**, 246–251 (2022).
58. Zhang, Y. et al. High-dimensional quantum key distribution by a spin-orbit microlaser. Physical Review X **15**, 011024 (2025).

## Acknowledgement

This work is supported by National Natural Science Foundation of China (62175076, 62105028, 62125503, 62261160388), Natural Science Foundation of Hubei Province of China (2023AFA028), and Hubei Optical Fundamental Research Center (HBO2025TQ004).

## Author contributions

S.Z. and M.M.Z. conceived the idea, J.L. and S.Z. designed the device and the chip layout, T.G.W. and J.L. carried out the chip characterization and far-field experimental measurements. J.L., S.Z., and K.Y.W. performed the simulation of inverse design. T.G.W. carried out the vector beam experimental measurements. S.Z., J.L., Z.Y.W, Y.M, Y.J.S. and M.M.Z. analyzed the experimental data, S.Z., J.W. and M.M.Z. supervised the project, S.Z., J.L. and M.M.Z. wrote the manuscript with the support from all co-authors.

## Competing interests

The authors declare no competing interests.